\documentclass[%
 reprint,
 amsmath,amssymb,
 aps,
]{revtex4-1}

\usepackage{graphicx}
\usepackage{dcolumn}
\usepackage{bm}
\usepackage{amssymb}

\begin{document}


\title{Simulating the Ising Model with a Deep Convolutional \\ Generative Adversarial Network}

\author{Zhaocheng Liu}
 \affiliation{%
 School of Electrical and Computer Engineering,\\ Georgia Institute of Technology, Atlanta, Georgia 30332, USA
}%
\author{Sean P. Rodrigues}%
 \affiliation{%
 School of Electrical and Computer Engineering,\\ Georgia Institute of Technology, Atlanta, Georgia 30332, USA
}%
\author{Wenshan Cai}
 \email{wcai@gatech.edu}
 \affiliation{%
 School of Electrical and Computer Engineering,\\ Georgia Institute of Technology, Atlanta, Georgia 30332, USA
}%
 \affiliation{%
 School of Materials Science and Engineering, \\ Georgia Institute of Technology, Atlanta, Georgia 30332, USA
}%


\begin{abstract}
The deep learning framework is witnessing expansive growth into diverse applications such as biological systems, human cognition, robotics, and the social sciences, thanks to its immense ability to extract essential features from complicated systems. In particular, recent developments of the field have revealed the unique faculty of deep learning to accurately approximate complex physical systems in fluid dynamics, condensed matter physics, etc. The convolutional neural network (CNN) is an efficient approach to represent complex systems with large degrees of freedom. On the other hand, the generative adversarial network (GAN), as an unsupervised learning algorithm, is capable of efficiently imitating the distribution of training data. Here we leverage this unique property of GAN, in conjunction with CNN methodology, to establish an Ising simulator whose generator can produce Ising states given temperature $T$ around the criticality. The generated Ising states well resemble, and essentially replicate, the data from conventional Monte Carlo simulations. Our results demonstrate the universality of GAN as a promising tool in the field of computational and statistical physics.
\end{abstract}

\maketitle


\section{Introduction}
As degrees of freedom in complex systems increase, accurately replicating system behavior has become an ever escalating challenge for the scientific community. However, with the rise of the big data era, vast amounts of information can be recorded, allowing the use of data driven methods to efficiently approximate these complex systems. Among these machine learning approaches, deep learning has shown the greatest potential in the study of complicated physical phenomena such as quantum physics, high energy physics, and astronomical sciences [1-6]. In the past few years, the deep learning framework has been adopted for diverse branches of the physical sciences.

From the perspective of information theory, the success of deep learning is, in part, attributed to its back-propagation training method and efficient way to reduce the dimension of the system under consideration [7,8]. The restrict Boltzmann machine (RBM), as a special case of the deep learning architecture, represents an exact mapping of renormalization group theory [9,10]. In addition, it has been shown that RBM can be adopted as a neural network state to precisely describe quantum entanglement in many-body systems [11,12]. As a result, many-body problems, such as the approximation of ground states of a quantum system [13], the detection of features of quantum entanglement [14], and the acceleration of traditional quantum Monte Carlo (MC) algorithms [15], can now be tackled efficiently when deep learning methods are utilized in conjunction with many-body theory.

Many body systems have also been successfully described by using the convolutional neural network (CNN), a technique originating from computer vision [4]. In particular, networks involving the CNN architecture can be used to predict phase transitions and calculate system Hamiltonians under supervised learning frameworks [16-19]. While the reasoning behind CNN’s ability to make such accurate approximations for complex physical systems remains nebulous, the spatial translation symmetry in many-body systems is a necessary condition for the success of CNN [7].

On the other hand, generative models have been exploited to solve complex tasks in the machine learning community.  A wide variety of natural phenomena and human behaviors can be simulated through various unsupervised learning methods, such as the hidden Markov model, variational autoencoders and reinforcement learning [20-23]. In contrast to discriminative models that rely on the extraction of useful information, generative models produce information based on the characteristics of a system. Among all generative models, generative adversarial network (GAN) and its varieties are among the most studied models, thanks to their ability to simulate the distribution of training data in a zero-sum game framework [24-28]. This unique feature allows us to construct GAN models to imitate real, stochastic, physical systems.

In this work, we introduce an Ising simulator built by incorporating CNN into a GAN framework to simulate the Ising model near criticality. Given a specific temperature, the trained simulator is able to efficiently generate Ising states and faithfully replicate the results obtained from a conventional MC simulation. We will first review the Ising model and the convolutional neural network (CNN) in the general form, and will subsequently discuss the use of the deep CNN architecture for the representation of the two dimensional (2D) Ising model. In the central part of the work, we will develop a generative adversarial network (GAN) incorporated with CNN to simulate the 2D Ising model near the critical temperature, and verify the effectiveness of our approach by comparing the generated data  to the that from the MC simulation.

\section{Ising model and convolutional neural network}

In statistical physics, the Ising model is described as an ensemble of binary spins with coupling interactions in some lattices. Consider $N$ spins $s = \{ {s_i}\} $ that can take a value $ \pm 1$, where the index \emph{i} labels the positions of spin $s_i$ , then the Hamiltonian of the system together with the coupling constants $\{ {K_i}\} $  takes the form

\begin{equation}
H \left( s \right) =  - \sum\limits_i {{K_i}{s_i}}  - \sum\limits_{ij} {{K_{ij}}{s_i}{s_j}}  - \sum\limits_{ijk} {{K_{ijk}}{s_i}{s_j}{s_k}}  -  \ldots \,,
\end{equation}

\noindent where each summation is taken over the nearest spins. The probability of a spin configuration at temperature $T$ is given by the Boltzmann distribution 

\begin{equation}
P\left( {s|T} \right) = {e^{ - H\left( s \right)/kT}}/Z \,,
\end{equation}

\noindent where $k$ is the Boltzmann constant, and $Z$ represents the partition function defined as $Z = \sum\nolimits_s {P(s|T)} $ for the classical Ising model.  For simplicity and convenience, we rearrange all spins in the lattice and denote them as a state vector $\psi  = \left| {{s_1}{s_2} \cdot  \cdot  \cdot {s_N}} \right\rangle $  in the following discussion.

To illustrate the utility of CNN for the representation of the Ising model, we start by investigating a convolutional operation on each Ising state $\psi $ with $N$ spins. Consider a 2D Ising model shown in Figure 1a, here we define a kernel $h \in {\mathbb{R}^{n \times n}}$ applied to a given state. Each time we multiply the spin states in that kernel with the corresponding entry of the kernel in a point-by-point manner, and sum the $n \times n$ results into a new value. This multiplication process is carried over all spins in the lattice with shift (or stride) to reach a new state $\psi '$. It is obvious that the transformed state possesses a number of spins $N'$ less or equal to $N$. For instance, in the case shown in Fig. 1a with a stride number of 2, we transform a 36-spin state to a 9-spin state (or a 6-spin state if the rightmost and bottom spins are included). We can express this operation as $\psi ' = h * \psi $, or in a matrix representation:

\begin{equation}
\psi ' = P\psi \,,
\end{equation}

\noindent where $P \in {\mathbb{R}^{N' \times N}}$. The resultant state $\psi '$ after the convolution has each of its units encoded with the interaction information of the spins in $\psi $. We can apply $M$ different kernels to a state in a cascaded manner, so that $\psi ' = {P_M}{P_{M - 1}} \cdot  \cdot  \cdot {P_1}\psi $, which results in a state $\psi '$ with dimension much less than the original $\psi $. This cascaded process, with loss of information, serves as a measure to the state $\psi $, while the state $\psi '$ contains the value of this sequential measure.

\begin{figure}
\centerline{\includegraphics[width=9cm]{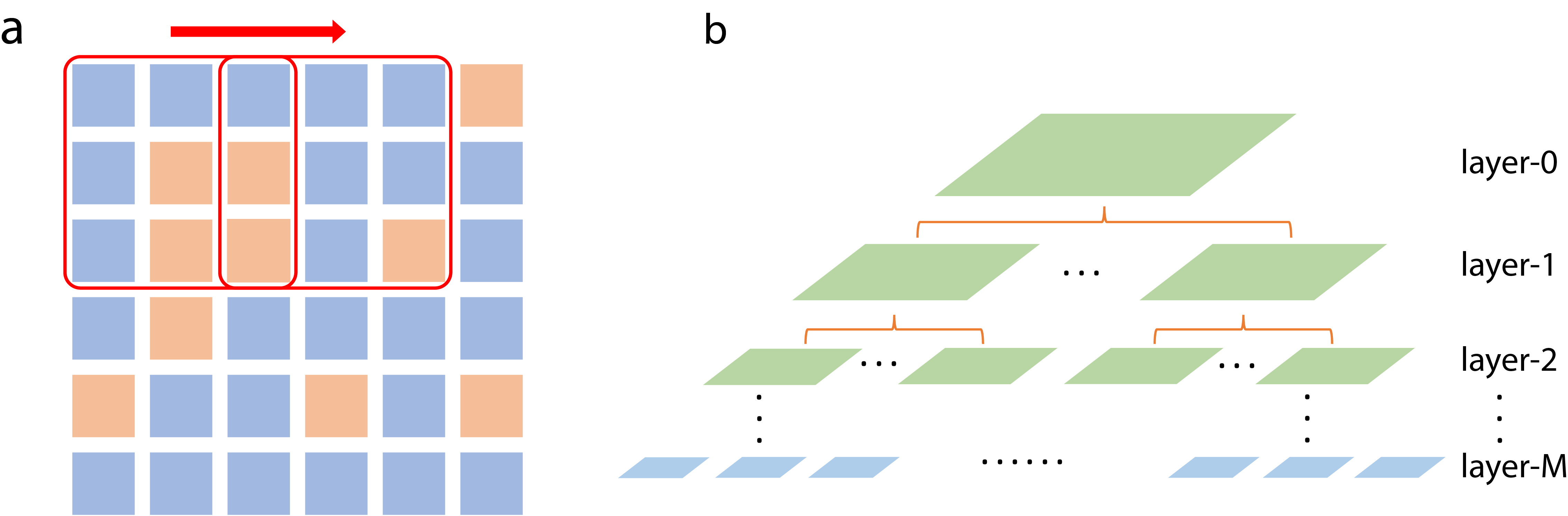}}
\caption{Schematic of convolutional neural network transformations. (a) Schematic of the convolutional operation by a kernel $h \in  \mathbb{R} ^ {3 \times 3} $  on a 2D square lattice with a stride of 2. (b) State transformation using a convolutional neural network. States in each layer are derived from its preceding layer, based on a set of convolutional operations and nonlinear activations. The final layer contains a collection of states with dimensions much lower than the original one. }
\end{figure}

Although the cascaded process described above reduces the system dimension by encoding the coupling information, it is clear that a one-measure process contains too limited information to describe the entire system. The architecture of CNN provides us with an efficient approach to describe the state $\psi $ in many ways. The basic structure of CNN contains convolutional layers, which are followed by nonlinear layers (activation layers). The transformation of states through an $M$-layer CNN is illustrated in Fig. 1b. Suppose there is a single state, ${\psi ^0}$, at layer-0 ${L_0}$, after applying ${N_k}$ different kernels to ${\psi ^0}$, we obtain ${N_k}$ separate states without nonlinear activations, which can be denoted as a set $\psi '$. We then apply a nonlinear activation function to each of the elements in $\psi '$, resulting in $\psi '$ transformed to ${\psi ^1}$ in layer-1  ${L_1}$. Repeating this two-step process of convolutional operation and nonlinear activation to each state in every layer, as shown in Fig. 1b, results in a set of states with lower dimensions.

In the next step we aim to develop an explicit expression of this transformation. First we denote layer-$i$ as ${L_i}$, the number of states in ${L_i}$ as ${N_{{S_i}}}$, the number of kernels after ${L_i}$ as ${N_{{k_{i + 1}}}}$, and the $k$-th kernel after ${L_i}$ as $h_k^{i + 1}$, where $1 \le k \le {N_{{k_{i + 1}}}}$. In ${L_i}$, we concatenate the states $\left\{ {\psi _j^i} \right\}_{j = 1}^{{N_{{S_i}}}}$ using direct sum ${\psi ^i} =  \oplus _{j = 1}^{{N_{{S_i}}}}\psi _j^i$.  After applying kernel $h_k^{i + 1}$ to ${\psi ^i}$, we obtain one of the states at ${L_{i+1}}$ as $\psi _k^{i + 1} = \left( {{ \oplus ^{{N_{{S_i}}}}}P_k^{i + 1}} \right){\psi ^i}$, where $P_k^{i + 1}$ stands for the corresponding matrix representation of $h_k^{i + 1}$.  We then define

\begin{equation}
{P^i} = \left[ {\begin{array}{*{20}{c}}
{{ \oplus ^{{N_{{S_{i - 1}}}}}}P_1^i}\\
{{ \oplus ^{{N_{{S_{i - 1}}}}}}P_2^i}\\
 \vdots \\
{{ \oplus ^{{N_{{S_{i - 1}}}}}}P_{{N_{{k_i}}}}^i}
\end{array}} \right] .
\end{equation}

\noindent Without nonlinear activation, the state ${\psi ^{i + 1}} =  \oplus _{k = 1}^{{N_{{k_{i + 1}}}}}\psi _k^{i + 1}$ in ${L_{i+1}}$, can be written in a compact form

\begin{equation}
{\psi ^{i + 1}} = {P^{i + 1}}{\psi ^i} \,,
\end{equation}

\noindent which is a linear transformation of ${\psi ^i}$.  Taking into account the nonlinear activations, Eq. (5) can be rewritten as ${\psi ^{i + 1}} = {\tilde P^{i + 1}}\left( {{\psi ^i}} \right)$,  where ${\tilde P^{i + 1}}$ is a composition of nonlinear mapping and convolutional operations. Furthermore, considering the full architecture of the $M$-layer CNN in Fig. 1b, we represent the transformation as ${\bf{P}} = {\tilde P_M} \circ {\tilde P_{M - 1}} \cdot  \cdot  \cdot  \circ {\tilde P_1}$, such that the final state after CNN becomes

\begin{equation}
{\psi ^f} = {\bf{P}}\left( {{\psi ^0}} \right) .
\end{equation}

\noindent ${\psi ^f}$ resides in a nonlinearly transformed space with dimension lower than ${\psi ^0}$, and each element of ${\psi ^f}$ is a representation of the original system, but with cascaded measures each defined by different kernels. Thus, the space in which ${\psi ^f}$ resides is a phase space of the original system. The training process of CNN by backpropagation is to seek the best set of kernels, so that the CNN maps the original Ising states with similar properties into close regions in phase space. CNN is superior to fully connected network because of the unique feature of its transformations. These transformations can be afforded thanks to the translational symmetry of the Ising model, which allows CNN to store the local coupling information throughout the process, while maintaining small amounts of parameters of the network. This intriguing feature of CNN speeds up the training process, but preserves the expressiveness of the network. This is particularly important when a gigantic amount of measures are needed and thereby the network is deep. The description of the system by phase space, facilitates the tackling of classification and regression problems by using additional networks and machine learning algorithms. With this strategy in place, the pivotal job now is to adopt the best optimization method to train the CNN with respect to a specific training aim.

\section{Ising simulator with a convolutional generative adversarial network}

Based on the analysis above, we will build a generative adversarial network (GAN) with CNN to simulate the  Ising model at a given temperature $T$. In other words, we train a generator in the GAN to produce Ising states ${\psi _G}$, whose distribution ${P_G} = P\left( {{\psi _G}|T} \right)$ approximates the Boltzmann distribution given by ${P_B} = P\left( {\psi |T} \right)$, where $\psi $ is the true state configuration. In this network, we use a variation of CNN called transpose convolutional neural network (TCNN), which functions inversely to CNN, i.e., TCNN transforms states in phase space into Ising states. Mathematically, the transpose convolutional operation in TCNN is exactly the transpose of convolutional operation matrix in CNN. As a result, TCNN is also capable of representing the Ising model efficiently, but in a bottom-up manner as indicated in Fig. 1b. 

\begin{figure}
\centerline{\includegraphics[width=9cm]{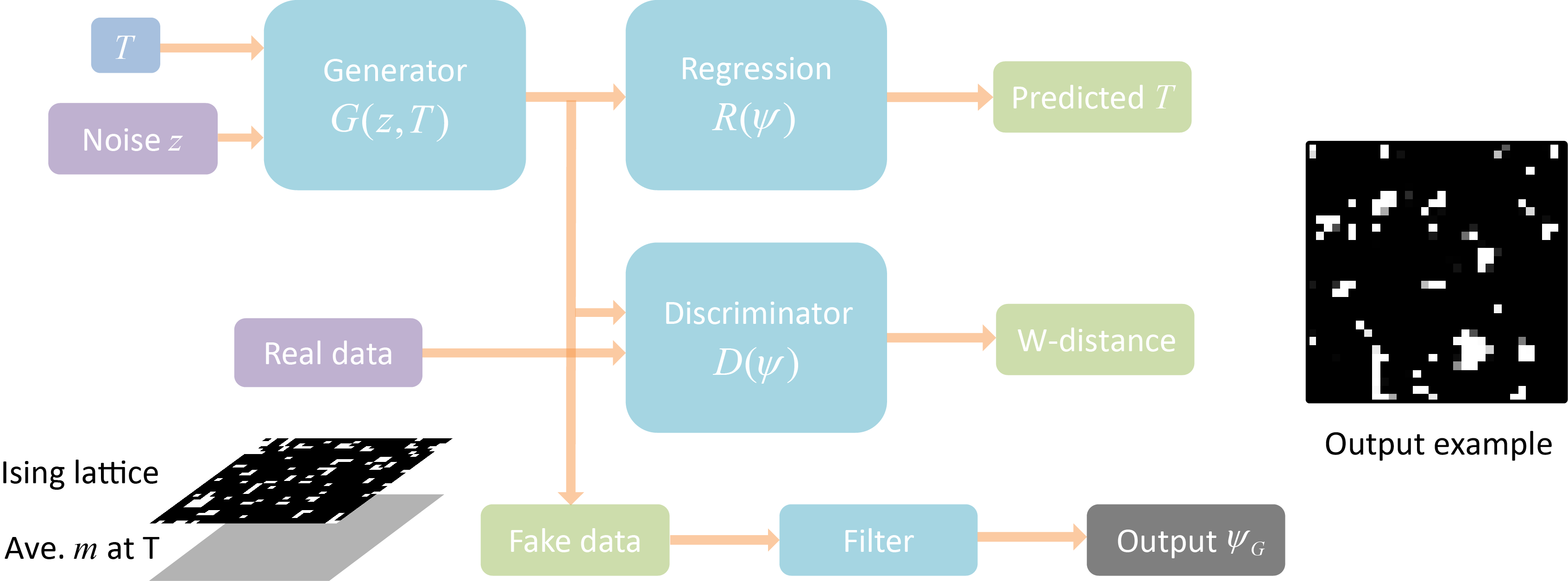}}
\caption{Architecture of a generative adversarial network for the Ising simulator. Two CNNs serve as the discriminator and the regression network, respectively, and a transpose CNN acts as the generator. All the CNNs in this framework share the same structure. The final output states are filtered from the “fake” data derived from the generator. The training data are augmented by adding a second layer of magnetization information at the given temperature $T$, in order to map the originally overlapped distributions into other dimensions.}
\end{figure}

The whole structure of our network is illustrated in Figure 2. Three CNNs, with the same architecture as defined in [28], are used as a generator $G\left( {z,T} \right)$, a discriminator $D\left( \psi  \right)$ and a regression network $R\left( {{\psi _G}} \right)$. The standard GAN contains two networks – a generator $G\left( {z,T} \right)$ and a discriminator $D\left( \psi  \right)$, which are TCNN and CNN respectively in our case. The generator takes random noise $z$ as a set of inputs and generates a “fake” state ${\psi _G}$, and the discriminator takes both real samples and samples from the generator as inputs. The output of the discriminator determines the distance of distribution between the real samples and the generated ones. Here we choose Wasserstein distance (W-distance) as the discriminator output introduced by Arjovsky \emph{et al.} [26], which is known as an accurate metric to evaluate the  distance between the generated distribution and its real counterpart. Other than standard GAN, we also need to train the generator so that it produces proper states based on the input temperature $T$. For this purpose, we define a regression network $R\left( {{\psi _G}} \right)$ in order to predict the temperature $\hat T $ of the generated Ising state and optimize the generator so that the predicted temperature $\hat T$ is close to the given temperature $T$.

The training details are as follows. We first run through the algorithm which is used to optimize the GAN with the Wasserstein metric to train $D$ and $G$. In this process, we first take some of the real states ${\psi}$ and the generated ones ${\psi _G}$ from random input $T$, and then feed them into the discriminator. The loss of the discriminator is defined as the approximated W-distance of the generated samples ${\psi _G} = G(z,T)$ and real ones $\psi $:

\begin{equation}
{L_D} = \mathbb{E} \left[ {D\left( {{\psi _G}} \right)} \right] - \mathbb{E} \left[ {D\left( \psi  \right)} \right] .
\end{equation}

\noindent The expectation $\mathbb{E}$ is taken over all samples at each training iteration. Meanwhile, the generator is updated by using the output of the discriminator $D$:

\begin{equation}
L_G^1 =  - \mathbb{E} \left[ {D(G(z,T))} \right].
\end{equation}

\noindent After several iterations of the training process described above, we update the generator by minimizing the difference of the temperature $\hat T$  to the input $T$. The temperature $\hat T$ can be estimated by the regression network $\hat T = R\left( {{\psi _G}} \right)$. Hence, we add a second loss of the generator:

\begin{equation}
L_G^2 = \mathbb{E} \left[ {{{\left| {T - \hat T} \right|}^2}} \right].
\end{equation}

\noindent For efficiency considerations, we pretrain the regression network R until the relative error of its estimation reaches below $1 \times 10 ^ {-5}$.

A major problem of the training scheme outlined above is that the distributions for two given temperatures may overlap substantially when the temperatures are close enough. This means the discriminator and the regression network are not able to judge the temperature merely based on the Ising state $\psi $. To distinguish the distribution at a continuous temperature $T$, we define an auxiliary state ${\psi ^a}$ for each $\psi $, which encodes additional information related to $T$. This auxiliary state ${\psi ^a}$ is to map the distribution to other dimensions. As a result, only ${\psi ^a}$ needs to be a function of $T$ such that a distinct temperature, ${T_1} \ne {T_2}$ immediately leads to a difference in the state ${\psi ^a}({T_1}) \ne {\psi ^a}({T_2})$. A straightforward choice of ${\psi ^a}$ for the Ising model is the average magnetization $m(T)$ at $T$. Additional information, such as the average energy or any self-defined function, can also be encoded and potentially improve the performance, albeit the cost of slower process of the data processing and training. With all these factors in consideration, our final input of the real state and the generated state will be the concatenation of the Ising states and the auxiliary states $\tilde \psi  = \psi  \oplus {\psi ^a}$. In practice, we construct ${\psi ^a}$ with the same size as ${\psi}$, where ${\psi ^a}$ contains a uniform value given by $m(T)$.

Since GAN with the W-distance metric has sufficient capability to generate $\tilde \psi $ as a whole, fixing the auxiliary state ${\psi ^a}$ will enforce the Ising state ${\psi}$ to adjust itself. The auxiliary states further help to train the generator $G$ to generate Ising states with conditional temperature information. We define the loss:

\begin{equation}
L_G^3 = \mathbb{E} \left[ {{{\left\| {{\psi ^a} - \hat \psi _G^a} \right\|}_2}} \right] ,
\end{equation}

\noindent where ${\psi ^a}$ is the auxiliary state of the output from the generator. This loss is constructed to minimize the distance between the auxiliary states of the real and generated samples, and consequently to enforce the generator to produce states at the given input $T$. 

The total loss of the generator is given by the combination of the three losses with penalties ${\lambda _1}$ and ${\lambda _2}$:

\begin{equation}
{L_G} = L_G^1 + {\lambda _1}L_G^2 + {\lambda _2}L_G^3 .
\end{equation}

\noindent where ${\lambda _1}$ and ${\lambda _2}$ are taken to be 100 in our case. However, as the losses $L_G^2$ and $L_G^3$ are not fully compatible with the W-distance loss $L_G^1$, the generator may generate Ising states with temperatures away from the input $T$. To resolve this issue, we set a filter to eliminate this portion of states. In practice, we compute the average value of auxiliary states $\hat m = \bar \psi _G^a\left( {z,T} \right)$, set a window size $w$, and only accept the generated states that satisfy $\left| {m(T) - \hat m} \right| \le w/2$.

\section{Simulation results and analysis}

\begin{figure}
\centerline{\includegraphics[width=9cm]{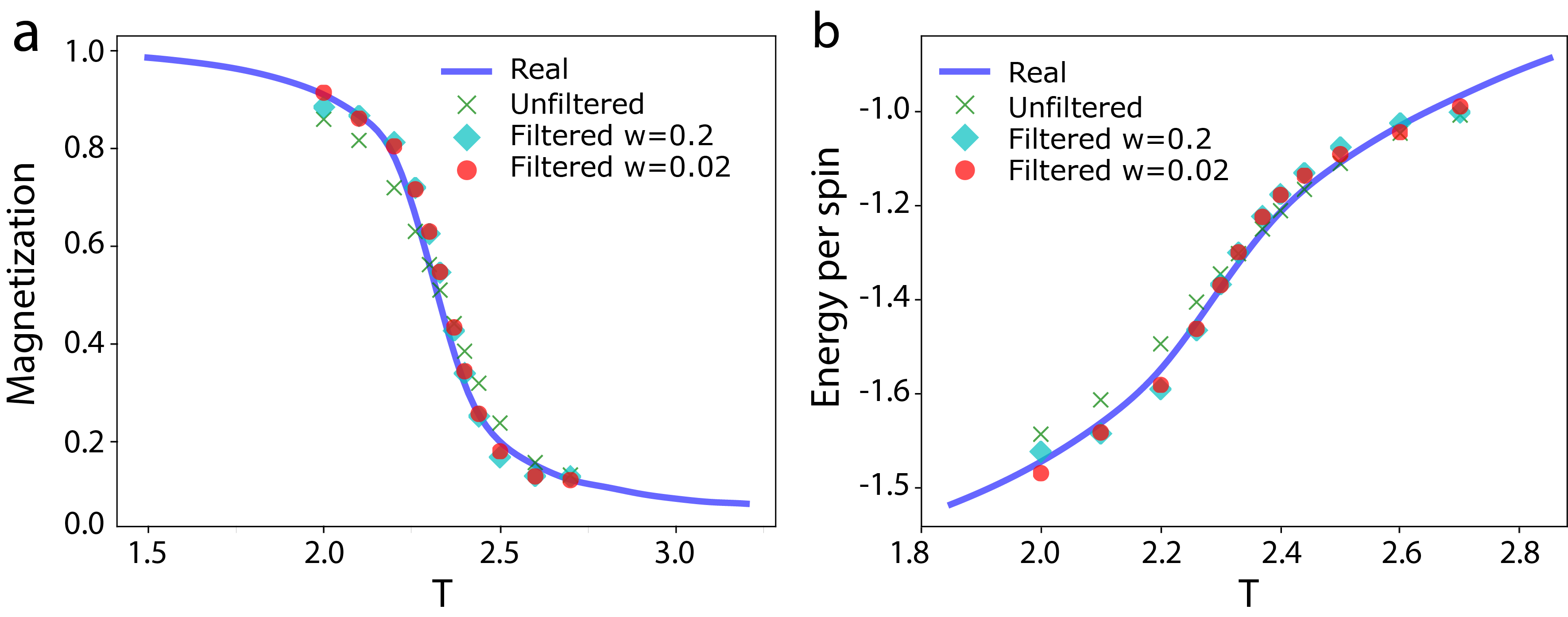}}
\caption{Average magnetization and energy per spin of the trained Ising simulator. (a) Expectation of magnetization of the generated Ising states from $T = 2.0$ to $T = 2.7$, with and without filters. (b) Expectation of energy per spin of the generated Ising states for the same temperature range and filter settings as in (a). The generated states progressively approach the accurate representation obtained from the MC simulation (solid blue curve) when the window size of the filter shrinks.}
\end{figure}

To verify the effectiveness of the network described above, we build an Ising simulator that produces Ising states in a $32 \times 32$ lattice near its critical temperature. The model we are simulating is descried by a Hamiltonian of the form

\begin{equation}
H =  - J\sum\limits_{ < i,j > } {{s_i}{s_j}} ,
\end{equation}

\noindent where $J$ represents a coupling constant, and the summation is taken over all nearest neighbors. We define $J = 1$ in the following simulation for simplicity. The training data of the Ising states are generated directly from a Monte Carlo clustering algorithm from $T = 2.0$ to $T = 2.7$ with 1000 states at each 0.1 increment and 100 at each 0.01 increment. The states are sampled for each of the 50 iterations in order to minimize the correlation of adjacently sampled states. The noise is set as 100 uniformly random variables from $-1$ to $1$. After the training process, we extract the generator and the filter to form our Ising simulator. When testing the simulator, we generate 10,000 Ising states for each target temperature $T$, and compare the distribution of the generated states to that of the states sampled from the MC simulation.

The comparison of the generated data from our Ising simulator, including the average magnetization and the energy per spin of the output states, to the real data derived from the MC simulation is shown in Figure 3. The data from our generator without filter well resembles the trend obtained from the MC simulation. The deviations observed in the figure stem from the generated states with temperatures much lower or higher than the given $T$. To suppress this deviation, we implement additional filters around the true value of $m(T)$ described before. From an accuracy perspective, the smaller the size $w$ is, the more precise the temperature of the generated state would be. This is better visualized through the two results in Fig. 3, with filter window of $w = 0.2$ and $w = 0.02$, respectively. However, a very small w would incur an increased amount of rejections of the generated states with temperature reasonably close to real data, and this would adversely slow down the generating process. Therefore, unless otherwise specified, we choose $w = 0.2$ for all the cases in the following discussion.

\begin{figure}
\centerline{\includegraphics[width=9cm]{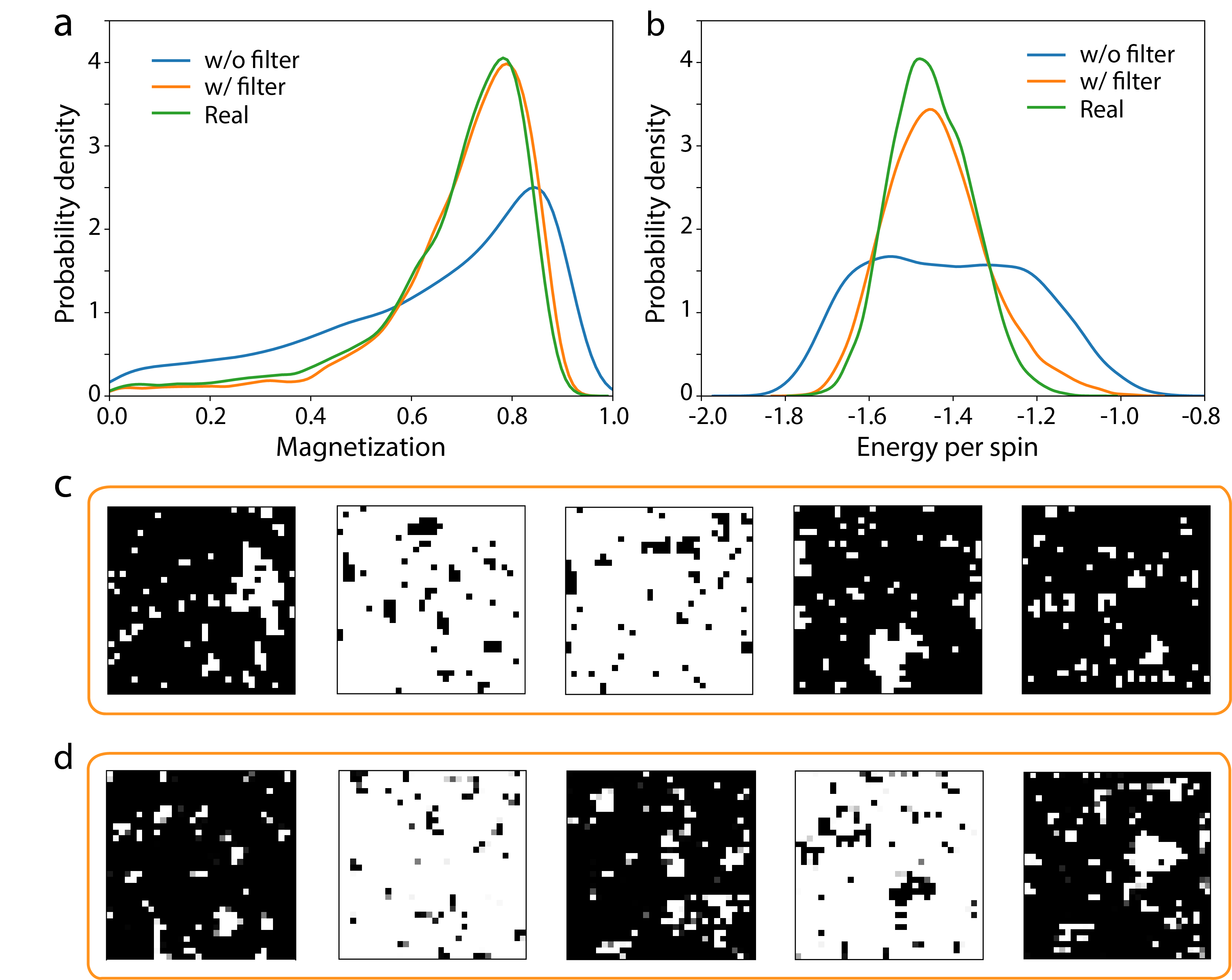}}
\caption{Distributions of the output from the trained Ising simulator at $T = 2.26$. (a) The probability density distribution of magnetization with (orange) and without (blue) the filter, along with the real data obtained from the MC simulation (green) at $T = 2.26$. (b) The probability density distribution of energy per spin with the same temperature and filter settings as in (a). (c) The Ising states randomly sampled using the MC algorithm at the specified temperature. (d) Ising states generated from our GAN Ising simulator. The values of the output states are not manually rounded to $\pm 1$, leaving the intermediate values in grayscale. }
\end{figure}

To further illustrate the strength of our scheme, we test the distributions of the magnetization and energy of the generated Ising states at a temperature of $T = 2.26$, which is marginally higher than the critical temperature. As shown in Figure 4a and 4b, the distribution of the generated states without the filter (blue lines) exhibits a noticeable deviation compared with the real distribution (green lines), although the expectations of them are approximately the same according to Fig. 3. This observation indicates that the network masters some statistical behavior of the Ising model, but still generates states whose temperatures are far from the actual $T$. After applying the filter (blue lines), the output distribution fits remarkably well with the data generated from the MC simulation. Figures 4c and 4d display the randomly sampled states obtained from our Ising simulator as compared against the MC simulation at a temperature of $T = 2.26$. As a way to demonstrate the robustness of our model, we choose not to round the output to $\pm 1$, and instead leave the intermediate values in grayscale (Fig. 4d).  From the comparison, we conclude that our deep-learning based Ising simulator successfully masters the topological characteristics of the true states without losing details. It is worth noting that the GAN learns the distribution of all states rather than just memorizing the state of the training samples. In other words, the outputs from the GAN are not simply reconstructions of the learning samples; instead, they are totally random states with respect to the input noise.

\begin{figure*}
\centerline{\includegraphics[width=16cm]{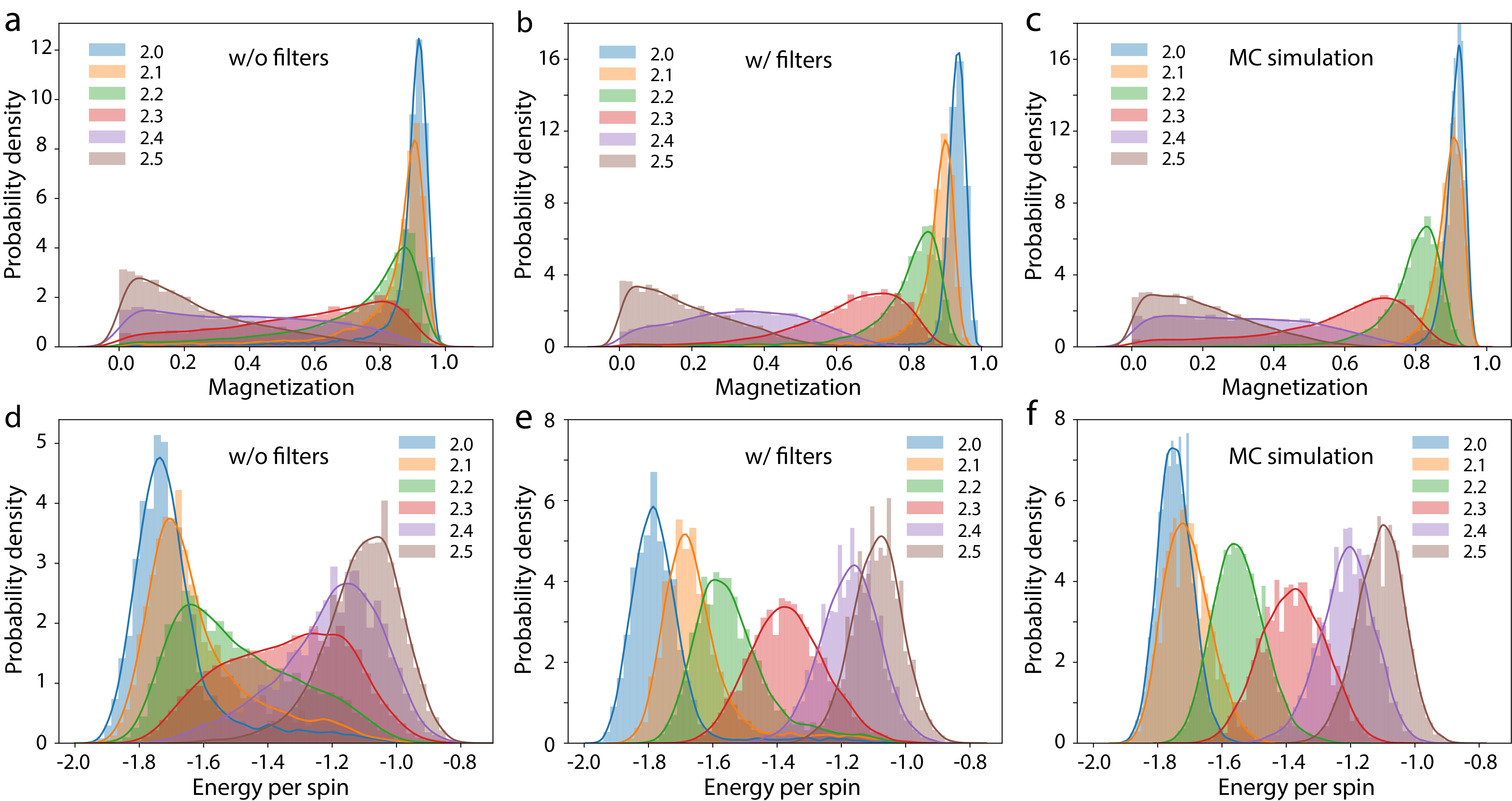}}
\caption{Statistical distributions of the output from the Ising simulator near criticality.  (a-c) Distributions of magnetization of the Ising states sampled from the GAN simulator without (a) and with (b) the filter, plotted against the MC solution (c). The sampling temperature is set from $T = 2.0$ to $T = 2.5$ with intervals of 0.1.  (d-f) Distributions of the energy per spin of the Ising states for the same conditions used in (a-c). The filter size is set to $w = 0.2$ for all temperatures, except for the lowest temperature of $T = 2.0$, at which a narrower window of $w = 0.02$ is used to better distinguish the values in the second layer of the output.}
\end{figure*}

Finally, we show in Figure 5 the histograms of the magnetization and energy obtained from our deep-learning simulator and the MC algorithm, respectively, for temperatures from 2.0 to 2.5 at 0.1 intervals. For distributions without filtering (Figs. 5a and 5d), the overall shape of the distribution spreads out from the actual distribution derived from the MC simulation shown in Fig. 5c and 5f. When the filter is implemented (Figs. 5b and 5e), the distributions of the magnetization and energy essentially replicate the actual distributions with negligible deviations.

\section{Conclusion}
In summary, we have demonstrated the feasibility to approximate and simulate the Ising model in the deep learning framework with remarkable accuracy. Consolidating the exceptional capacity of CNN to represent complicated many-body systems, we have developed an effective scheme to train an Ising simulator by employing a deep convolutional generative adversarial network. The Ising states generated from our simulator successfully replicate the distributions obtained from the traditional Monte Carlo simulation. Thanks to the unique nature of the GAN, the Ising states generated from our simulator are not simply reconstructions of the existing ones, but states with entirely new configurations determined by input noise. The model described in this work represents a prototype for simulating the 2D Ising model with deep neural networks, whose precision can be further improved by adopting more sophisticated training methods. We envision the broad utility of such generative networks for simulations of general many-body systems, with diverse potential applications such as the acceleration of traditional sampling algorithms, the approximation of large-scale molecular dynamics, and the accurate representation of complex systems based on experimental observations.

\bibliography{}
\noindent[1]	Y. LeCun, Y. Bengio, and G. Hinton, Nature \textbf{521}, 436 (2015).

\noindent[2]	P. Baldi, P. Sadowski, and D. Whiteson, Nat. Commun. \textbf{5}, 4308 (2014).

\noindent[3]	P. Baldi, P. Sadowski, and D. Whiteson, Phys. Rev. Lett. \textbf{114}, 111801 (2015).

\noindent[4]	J. Carrasquilla and R. G. Melko, Nat. Phys. \textbf{13}, 431 (2017).

\noindent[5]	F. Schindler, N. Regnault, and T. Neupert, Phys. Rev. B \textbf{95}, 245134 (2017).

\noindent[6]	Y. Chen, Z. Lin, X. Zhao, G. Wang, and Y. Gu, IEEE J. Sel. Top. Appl. \textbf{7}, 2094 (2014).

\noindent[7]	H. W. Lin, M. Tegmark, and D. Rolnick, J. Stat. Phys. \textbf{168}, 1223 (2017).

\noindent[8]	R. Shwartz-Ziv and N. Tishby, arXiv:1703.00810  (2017).

\noindent[9]	P. Mehta and D. J. Schwab, arXiv:1410.3831  (2014).

\noindent[10]	D. H. Ackley, G. E. Hinton, and T. J. Sejnowski, Congnitive Sci. \textbf{9}, 147 (1985).

\noindent[11]	G. Carleo and M. Troyer, Science \textbf{355}, 602 (2017).

\noindent[12]	X. Gao and L.-M. Duan, Nat. Commun.\textbf{ 8}, 662 (2017).

\noindent[13]	A. Morningstar and R. G. Melko, arXiv:1708.04622  (2017).

\noindent[14]	D.-L. Deng, X. Li, and S. Das Sarma, Phys. Rev. X \textbf{7}, 021021 (2017).

\noindent[15]	L. Huang and L. Wang, Phys. Rev. B\textbf{ 95}, 035105 (2017).

\noindent[16]	P. Broecker, J. Carrasquilla, R. G. Melko, and S. Trebst, Sci. Rep.\textbf{ 7}, 8823 (2017).

\noindent[17]	Y. Zhang and E.-A. Kim, Phys. Rev. Lett. \textbf{118}, 216401 (2017).

\noindent[18]	K. T. Schütt, F. Arbabzadah, S. Chmiela, K. R. Müller, and A. Tkatchenko, Nat. Commun. \textbf{8}, 13890 (2017).

\noindent[19]	Y.-Z. You, Z. Yang, and X.-L. Qi, arXiv:1709.01223  (2017).

\noindent[20]	S. R. Eddy, Curr. Opin. Struc. Biol. \textbf{6}, 361 (1996).

\noindent[21]	T. Jaakkola and D. Haussler, in\textit{ Advances in Neural Information Processing Systems} (1999), pp. 487.

\noindent[22]	D. P. Kingma and M. Welling, arXiv:1312.6114  (2013).

\noindent[23]	M. L. Littman, in \textit{Proceedings of the Eleventh International Conference on Machine Learning} (1994), pp. 157.

\noindent[24]	I. Goodfellow, J. Pouget-Abadie, M. Mirza, B. Xu, D. Warde-Farley, S. Ozair, A. Courville, and Y. Bengio, in\textit{ Advances in Neural Information Processing Systems} (2014), pp. 2672.

\noindent[25]	X. Chen, Y. Duan, R. Houthooft, J. Schulman, I. Sutskever, and P. Abbeel, in\textit{ Advances in Neural Information Processing Systems} (2016), pp. 2172.

\noindent[26]	M. Arjovsky, S. Chintala, and L. Bottou, arXiv:1701.07875  (2017).

\noindent[27]	I. Gulrajani, F. Ahmed, M. Arjovsky, V. Dumoulin, and A. Courville, arXiv:1704.00028  (2017).

\noindent[28]	A. Radford, L. Metz, and S. Chintala, arXiv:1511.06434  (2015).

\end{document}